\documentclass[twocolumn,prb]{revtex4}

\usepackage{graphicx}
\usepackage{amssymb}

\begin{document}


\title{Nucleation and growth of single wall carbon nanotubes}

\author{F. Beuneu}

\affiliation{Laboratoire des Solides Irradi\'es, CNRS-CEA,
\'Ecole Polytechnique, F-91128 Palaiseau, France}

\begin{abstract}
The nucleation and growth of single wall carbon nanotubes
from a carbon-saturated catalytic particle surrounded by
a single sheet of graphene is described qualitatively
by using a very restricted
number of elementary processes, namely Stone-Wales defects
and carbon bi-interstitials. Energies of the different configurations
are estimated by using a Tersoff energy minimization scheme.
Such a description is compatible with a broad variety of size
or helicity of the tubes. Several mechanisms of growth of the embryos
are considered: one of them is made more favourable
when the tubes embryos are arranged in an hexagonal network
in the graphene plane. All the proposed mechanisms can be
indefinitely repeated for the growth of the nanotubes.
\end{abstract}

\maketitle

\section{Introduction}\label{sec:intro}

Since their discovery nearly fifteen years ago, single wall carbon nanotubes (SWNT)
have received considerable interest from scientists: they are quite simple nanoscopic
objects, with fascinating physical properties; moreover, their potential applications,
in the field of nanosciences and nanotechnology, are very promising.
But one major challenge is to control the growth of SWNTs, in particular concerning
their diameter and helicity. This is the reason why a lot of literature was recently
devoted to the understanding of the catalytic nucleation and growth of these tubes.

A typical situation is the high-temperature catalytic growth of SWNTs:
small metallic particles of Ni or Co are heated in the presence of carbon,
by arc discharge, laser heating or CVD, which causes growth of bundles of
SWNTs perpendicular to the surface of the particles. Maiti et al. \cite{maiti97}
suggest a general model which seems to get a broad agreement: the metallic
particles are oversaturated in carbon, and a graphene layer wraps
their surface. Embryos of SWNTs, looking like half-fullerenes, can form
on this layer, and subsequently grow from their foot.
These authors present molecular-dynamics computations, using a Tersoff-Brener
potential, on (11,3) tubes taken as examples.

More recently, other reports on the same subject were published.
Gavillet et al. \cite{gavillet01} presented a high-resolution transmission
electron microscopy study of such a root-growth mechanism, completed
by a computer-simulation work using quantum molecular dynamics.
The paper by Kanzow et al. \cite{kanzow01} is another example of a
growth model ``in which precipitated graphene sheets detach from the surface
of a liquid catalyst particle, forming fullerenelike caps''.
Gavillet et al. \cite{gavillet04} gave an interesting review of experimental
and theoretical results on SWNT nucleation and growth; in particular, they stated
that ``a natural process is to imagine that carbon atoms are
incorporated at the root or at the tip where `defects' necessarily
occur: heptagons at the root and pentagons at the tip and/or metal--carbon
bonds''; these authors also addressed the role of the catalyst.
Recently, Ding et al. \cite{ding04} presented molecular dynamics
calculations on very small iron particles oversaturated with carbon,
giving rise to very irregular SWNTs.

The purpose of the present paper is to propose qualitative ideas
towards a better understanding of the nucleation and growth of SWNTs.
We show that a very small number of elementary defect types are required
to build SWNTs from a graphene surface. We also address the case where
tubes grow inside a bundle.

\section{General ideas and tools}\label{sec:tools}

Our starting point consists in a graphene plane, on which an embryo
of SWNT will grow. In order to check the stability and energy of
the proposed defect geometries, we performed energy minimizations
by using the Tersoff model \cite{tersoff88}. We worked mainly with
a nearly square portion of the graphene plane of 240 carbon
atoms, with periodic boundary conditions in both directions of
the plane. We also considered the case of a network of embryos,
using for that purpose a diamond-shaped unit cell with suitable
periodic boundary conditions. Our energy computations do not
pretend to be very precise; they intend to give some indications
for comparing different possible configurations.

We consider three elementary defects in the graphene plane:

a) the Stone-Wales defect \cite{stone86} is the simplest possible
point defect, which consists in a $90^\circ$ rotation of a pair of
C atoms, with some rearrangement of the C-C bounds:
the net result is the transformation of four hexagons
into two heptagons and two pentagons. In our configuration,
it corresponds to an extra energy of 9.1 eV.
This estimate is quite large compared to more precise calculations
from \textit{ab initio} approaches \cite{zhao02,jensen02} which
give energy values for the Stone-Wales (SW) defect in the 5 to 6 eV range.
We use however the Tersoff potential, due to its high simplicity;
in spite of its lack of accuracy, we believe that it can be useful
for comparing different growth scenarii.

b) the bi-interstitial enables to add extra C atoms to the
graphene without generating dangling bonds (which would be the case
with single interstitials). Our model for the bi-interstitial consists
in adding atoms onto two opposite sides of an hexagon. The net result
is again two heptagons and two pentagons, with a different topology
compared to the SW defect. It corresponds to an extra
energy of 11.9 eV (this energy is taken as the difference with that
of the same number of atoms if they were in a perfect graphene sheet).

c) the last defect is the dislocation. It is composed of a
pentagon-heptagon pair \cite{lauginie97}, the Burgers vector (BV)
being perpendicular to the pentagon-heptagon axis. It is quite
amusing to note that, in the very different framework of grain
growth, Cahn and Padawer pointed out the existence
of this defect in a honeycomb network
many years ago \cite{cahn65}. Like any dislocation,
it is topologically impossible to create ex nihilo such a defect.
It can only be created as a pair of dislocations of opposite BV,
or (we shall give examples in the following) as a side effect of
the evolution of other defects. The energy of an isolated dislocation
is known to diverge logarithmically with distance in an infinite crystal;
however, in practice, such a divergence being slow, and for a finite size
of samples or a finite distance between dislocations, an energy value
estimate can often be given. Building two dislocations from
such a dislocation pair in our finite graphene sample, we estimate
the individual dislocation energy to be about 11 eV.

Elementary dislocation theory teaches that dislocations can glide,
along a line in the present 2D configuration: this line is parallel to
the Burgers vector, i.e. perpendicular to the pentagon-heptagon axis.
It is interesting to note that Fig.~2(a) and (b) of
Ref.~\cite{nardelli98} is an example of the creation and
glide of two such dislocations, evidenced in a numerical simulation.
In fact, it is quite straightforward that a dislocation can glide
by one polygon with a single SW made on one edge of the heptagon.
It is worth mentioning that like classical 3D dislocations,
two dislocations with opposite BV must attract themselves.

The other defects mentioned before, the SW and the bi-interstials,
can also glide: both these defects can be seen as made from two
dislocations with opposite BV, which can glide individually.

\section{Nucleation of an embryo}\label{sec:nucleation}

Our model lies on the fact that an embryo is nucleated in the graphene
plane, so that it can grow from the foot, perpendicularly to the
plane. The cap of the embryo looks like a half-sphere, which means,
as many authors remarked, that -- from Euler's theorem -- it contains
exactly 6 pentagons (in reality, topology only dictates that,
when the polygons are forced to be heptagons, hexagons or pentagons,
the difference between the number of heptagons and pentagons must be 6).
As a consequence, there must be exactly 6 heptagons around
the foot of the embryo.

\begin{figure}
\includegraphics{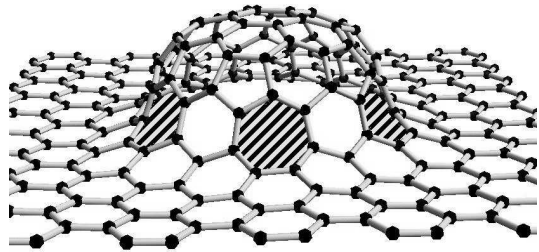}%
\caption{\label{fig:embryo} Embryo of a (12,0) zig-zag SWNT.
The heptagons in the foot of the embryo are hatched.
We get an extra energy of 28.1 eV.}
\end{figure}

We have constructed some embryos by adding bi-interstitials in
the gra\-phene plane. For instance, a (12,0) zig-zag embryo could
be built by adding 24 interstitials (12 bi-interstitials),
which generated 12 new polygons. Such an embryo is shown in
Fig.~\ref{fig:embryo}. The Tersoff energy minimization gives
an energy of 28.1 eV for this embryo.

\section{Growth}\label{sec:growth}

In order to make the (12,0) embryo grow one row, it is necessary
to add 12 hexagons, that is 24 interstitials. There are several
possibilities for adding these interstitials.

If the 12 bi-interstitials are added in the 12 polygons (6 hexagons
and 6 heptagons) which form the first ring of the embryo, the net
result is the growth of the tube without creating any supplementary
defect. The final energy of the tube is about 30 eV, which is only
slightly more than that of the embryo. Adding the bi-interstitials
not at once, but one after the other, the energy increases much more,
passing through a maximum of 9 eV above. It is clear that this
growing process can be repeated \textsl{ad libitum}, making
the tube grow indefinitely.

The fact that, in this process, interstitials are not added in the basal
graphene plane, but on the side of the embryo, can be questionable
if the C atoms come from the inside of the supersaturated metallic
particle. We wish thus to suggest here a second possible process.

\begin{figure}
\includegraphics{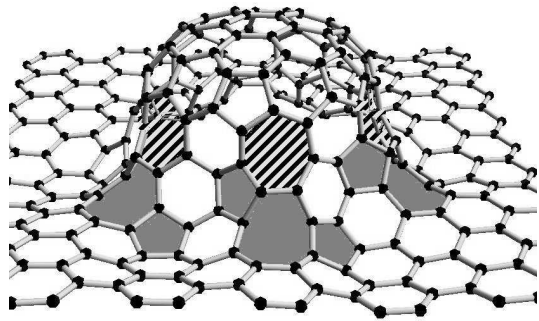}%
\caption{\label{fig:embryo24int} 24 carbon interstitials are added
on the foot of the embryo of a (12,0) SWNT. The heptagons in
the foot of the embryo are hatched. The pentagons and octagons
are grayed. We get an extra energy of 72.0 eV.}
\end{figure}

Adding the 12 bi-interstitials in the closest ring of hexagons
next to the foot of the embryo gives the arrangement shown in
Fig.~\ref{fig:embryo24int}, with an energy of 72.0 eV,
which is a much higher value; we discuss this value below.
Now, near the foot of the tube, a lot of defects are present:
6 octagons and 12 pentagons (they are grayed in
Fig.~\ref{fig:embryo24int}). These defects have to be eliminated
by some kind of glide movement in order to get a realistic
defect-free growing process. This can be done quite easily,
in two steps:

\begin{figure}
\includegraphics{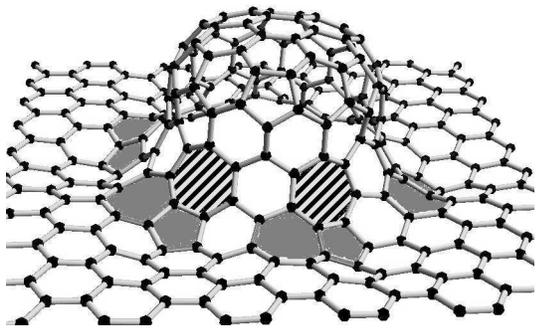}%
\caption{\label{fig:after6SW} 6 Stone-Wales transforms are made
on the pattern shown in Fig.~\ref{fig:embryo24int}.
The heptagons in the foot of the embryo are hatched.
The dislocations (pentagons plus heptagons) are grayed.
We get an extra energy of 52.7 eV.}
\end{figure}

a) the first one is required by the fact that the heptagons,
shown hatched in Fig.~\ref{fig:embryo24int}, have to go down
back to the foot of the tube. For doing that, 6 SW processes
on the bounds between heptagons and octagons are done, which
suppress also six of the pentagons and replace the 6 octagons
by 6 heptagons: this is clearly shown in Fig.~\ref{fig:after6SW}.
The energy is now 52.7 eV. The net result is a (12,0) tube,
grown by one row, plus, near its foot, 6 pentagon-heptagon pairs which are
6 dislocations, with the 6 possible BV values (the sum of these
BV is, of course, zero).

\begin{figure}
\includegraphics{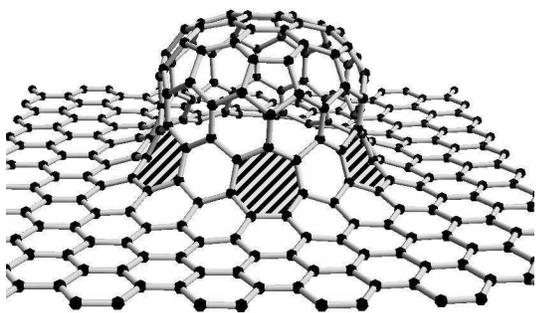}%
\caption{\label{fig:after12SW} 6 more Stone-Wales transforms are made
on the pattern shown in Fig.~\ref{fig:after6SW}. The embryo
of Fig.~\ref{fig:embryo} has grown of one row.
The heptagons in the foot of the embryo are hatched.
We get an extra energy of 30.4 eV.}
\end{figure}

b) the second step must be performed in order to annihilate these
dislocations. This can be done quite simply by 6 new SW processes,
which are done on the 6 bounds common to two heptagons. After doing
this, we are left with a perfect (12,0) tube, with an energy of
30.4 eV (Fig.~\ref{fig:after12SW}).

It is interesting to address more in detail the problem of the energy values,
which seem to be much higher in the second process described above.
Reality is more complex: Fig.~\ref{fig:embryo24int} to \ref{fig:after12SW}
show quite clearly what are the steps involved in the tube growth, but they
are certainly not the most economical path for this growth. It is much better
to do things like the following, from the embryo of Fig.~\ref{fig:embryo}:

--- introducing 4 interstitials near the foot of the embryo, giving birth to
one octagon and two pentagons.

--- making a first SW to transform these defects into one dislocation.

--- making a second SW to annihilate the dislocation.

--- repeating 5 more times these 3 steps, which gives finally the situation
depicted in Fig.~\ref{fig:after12SW}.

We have done this step-to-step process and monitored the energy, whose
maximum proved to be of the order of 16 eV above the energy of the perfect
embryo. This is a collar energy, substantially higher than
the one we have to pass in the first defect-free process
(however, the total energy is then about 44 eV, much lower than
that of the configuration in Fig.~\ref{fig:embryo24int}).
But the second process has the advantage of enabling the addition of
carbon interstitials in the graphene basal plane, not on the sides of
the embryo.

\begin{figure}
\includegraphics{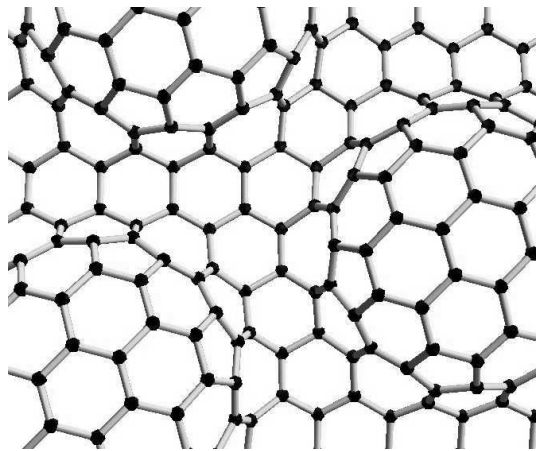}%
\caption{\label{fig:network} Embryos of (12,0) SWNTs, seen
from above, arranged in an hexagonal network. In this case,
the growth is possible without creation of dislocations (see text).}
\end{figure}

We finally tried to figure out what can be gained from considering
a network of embryos instead of an isolated one. We found that
when putting embryos in an suitable hexagonal network
(see Fig.~\ref{fig:network}), the second
process described above could be made a little simpler: after
addition of 24 interstitials per embryo, step a) described above
has to be done in order to bring down the heptagons. But step
b), with its 6 SW, is no longer needed: dislocations annihilate 3 by 3, between
neighbouring embryos. This simplification corresponds to an interesting gain
in energy, the collar value being about 10 eV above the embryo energy. This
is of the same order of magnitude as the first defect-free process.
We can also remark that this process of growth of SWNT bundles does not
require that all tubes have the same helicity: it is sufficient that the
geometry of the 6 created dislocations near the foot of each tube is the same,
which is much less restricting.

We point out that it would be interesting to get values for the energy barriers,
which the present calculations, taken at equilibrium, do not enable. However,
it remains difficult to understand how these growth processes involving very large
energies can be efficient in quasi-equilibrium processes. An interesting possibility,
considered by several authors \cite{gavillet04,jensen02}, is that catalytic effects
can reduce the energies involved.

\section{Conclusion}\label{sec:conclusion}

We have shown that the nucleation and growth of carbon nanotubes from
a graphene basal plane can
be qualitatively described with a small number of elementary
processes, each of them corresponding to a moderate amount of energy.
Such a description does not depend on the size or helicity of the tubes.
Several mechanisms can be considered: one of them is made more favourable
when the tubes embryos are arranged in an hexagonal network in the plane.
All the proposed mechanisms can be indefinitely repeated, giving rise to
a possible endless growth of the nanotubes.

\end{document}